\documentclass[seceq]{ptptex}

\usepackage{graphicx}
\usepackage{amsmath}

\newcommand{\ce}{\small\mathcal{E}}
\usepackage{bm}
\usepackage{braket}

\title{
An exact calculation of the transverse susceptibility for an antiferromagnetic Ising $\Delta$ chain %
}

\author{
Nobutaka \textsc{Kunisada}
and Yoshiyuki \textsc{Fukumoto}\footnote{E-mail: yfuku@ph.noda.tus.ac.jp}
}

\inst{
Department of Physics, Faculty of Science and Technology, \\
Tokyo University of Science, Noda, Chiba 278-8510
}

\abst{
We study the transverse susceptibility of the fully frustrated antiferromagnetic Ising $\Delta$-chain,
extending Minami's transfer-matrix method for the transverse susceptibility of general-type Ising linear-chains [\JPSJ{67,1998,2255}].   
For transverse fields $\Gamma_{1}$ on tip spin sites and $\Gamma_{2}$ on bottom spin sites, 
we calculate zero-field transverse-susceptibilities $\chi_{\rm{tip}}^{x}=\lim_{\Gamma_{1},\Gamma_{2}\to 0}M_{\rm{tip}}^{x}/\Gamma_{1}$ 
and $\chi_{\rm{bottom}}^{x}=\lim_{\Gamma_{1},\Gamma_{2}\to 0}M^{x}_{\rm{bottom}}/\Gamma_{2}$, 
where $M_{\rm{tip (bottom)}}^{x}$ denotes the magnetization for tip (bottom) spin sites. Both the transverse susceptibilities follow Curie's law at low temperatures. 
We also calculate $\chi_{\rm{bottom}}^{x}(\Gamma_{1}>0)$, transverse susceptibility of the bottom spin chain under finite tip-spin transverse-fields, to understand the Curie type behavior in the zero-field susceptibility.
Using the second-order perturbation theory, we discuss the $\Gamma_{1}$ dependence of $\chi_{\rm{bottom}}^{x}(\Gamma_{1})$ at zero temperature.
}

\begin{document}

\maketitle

\section{Introduction}
The tranverse-field Ising model has been often used as an effective model to study order-disorder transitions in some physical systems.\cite{rf:1} 
 There are many theoretical studies for this model. 
 In early time, de Gennes employed this model to understand ferro-electric ordering in Pottasium Dihydrogen Phosphate type systems. 
 As for the transverse-field Ising chain with only nearest neighbor interactions, 
 magnetic properties such as the free energy, magnetization, transverse susceptibility, etc, were obtained by Katsura, Pfeuty, and Fisher.\cite{rf:2,rf:3,rf:4,rf:5} 
 In those investigations, a mapping to a free fermion system via the Jordan-Wigner transformation was done, 
 and thus it is difficult to study more general models with $S\geq1$, second neighbor couplings, and so on. 

In 1990's, the exact transverse susceptibility for several type Ising chains was examined theoretically. 
Idogaki \textit{et al.} developed the differential operator method to obtain exact transverse susceptibility for a random-bond Ising chain.\cite{rf:6} 
Using this differential operator method, Kaneyoshi for the first time obtained the exact transverse susceptibility for mixed spin-$1/2$  and spin-$S$ ($S\geq1$) Ising chain.\cite{rf:7}

On the other hand, Minami used a transfer-matrix method to calculate the exact transverse susceptibility for the Ising linear-chain with arbitrary spin.\cite{rf:8} 
And later, he introduced more general Ising-type models with mixed spins and random bonds, 
and calculated the transverse susceptibility for these models by the transfer-matrix method.\cite{rf:9} 
Minami's results agree with the previous results obtained by Idogaki \textit{et al.} and Kaneyoshi, and he was also succeeded in getting some original and general results. 
According to Minami's procedure, the calculation of the exact transverse susceptibility is reduced to solving the eigenvalue problem of a finite order transfer matrix. 

\begin{figure}[t]
\centerline{\includegraphics[width=0.6\linewidth]{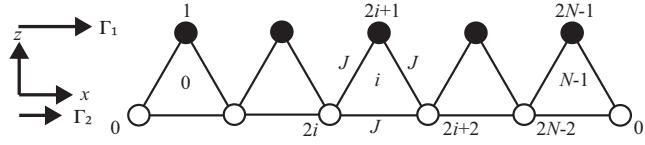}}
\caption{Geometry of the antiferromagnetic Ising $\Delta$-chain. 
The black circles denote tip spins and the white circles represent 
bottom spins. Note that couplings between bottom spins give fourth 
order terms of Fermi operators in the Hamiltonian when we use 
the Jordan-Wigner transformation.}
\label{fig:1}
\end{figure}

In this paper, we consider an exact calculation of the transverse susceptibility for an antiferromagnetic Ising $\Delta$-chain under transverse fields $\Gamma_{1}$ on tip spins and $\Gamma_{2}$ on the bottom spin chain  (see Fig.~\ref{fig:1}).
The Hamiltonian is written as
\begin{eqnarray}
   H=J\sum_{i=0}^{N-1}[s_{2i}^{z}s_{2i+2}^{z}+s_{2i+1}^{z}(s_{2i}^{z}+s_{2i+2}^{z})]
   -\Gamma_1Q_1-\Gamma_2Q_2,
\end{eqnarray}
where $\bm{s}_{i}$ are spin-$1/2$ operators, $N$ denotes the total number of triangles, and the periodic boundary condition,  $\bm{s}_{l+N}=\bm{s}_{l}$, is imposed. 
The operators $Q_1$ and $Q_2$, respectively, represent the magnetization operators for the tip and bottom spins defined by
\begin{eqnarray}
   Q_1=\sum_{i=0}^{N-1}s_{2i+1}^{x},\;\;\;\; Q_2=\sum_{i=0}^{N-1}s_{2i}^{x}.
\end{eqnarray}
We now define zero-field susceptibilities as follows:
\begin{eqnarray}
   \chi^{x}_{\rm{tip}}=\lim_{\Gamma_{1},\Gamma_{2}\to 0}M_{\rm{tip}}^{x}/\Gamma_{1},\;\;\;\;
   \chi^{x}_{\rm{bottom}}=\lim_{\Gamma_{1},\Gamma_{2}\to 0}M_{\rm{bottom}}^{x}/\Gamma_{2},
\end{eqnarray}
where $M_{\rm{tip}}^{x}=\langle Q_1 \rangle$ and $M_{\rm{bottom}}^{x}=\langle Q_2 \rangle$ are thermal averages of the magnetization operators under the Hamiltonian $H$.

 \begin{figure}[b]
\centerline{\includegraphics[width=0.8\linewidth]{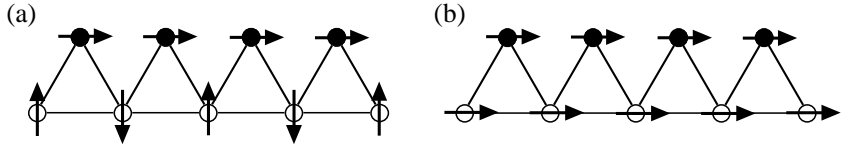}}
\caption{Schematic representations of (a) staggered state and 
(b) transverse-field paramagnetic state.}
\label{fig:1b}
\end{figure}

The present Ising $\Delta$-chain, without the transverse fields, is one of the simplest fully frustrated Ising models with a macroscopic ground state degeneracy,
and infinitesimal transverse fields determine the ground state uniquely.\cite{rf:10,rf:11} 
The transverse-field Ising $\Delta$-chain has been known to have a ground-state quantum phase transition as a function of the ratio of $\Gamma_{1}$ and $\Gamma_{2}$.\cite{rf:11} 
We set $\Gamma_{1}=\Gamma$ and $\Gamma_{2}=\lambda\Gamma$.
The staggered state defined in Fig.~\ref{fig:1b} (a), which involves spontaneous symmetry breaking in the bottom spin chain, is stabilized for $\lambda<\lambda_{\rm c}(\Gamma)$,
and the transverse-field paramagnetic phase defined in Fig.~\ref{fig:1b} (b) is stabilized for for $\lambda>\lambda_{\rm c}(\Gamma)$.
As $\Gamma$ is decresed, $\lambda_{\rm c}(\Gamma)$ is increased and $\lambda_{\rm c}(+0)=0.984$.

When an infinitesimal transverse field stabilizes the transverse-field paramagnetic phase, then $M_{\rm{bottom}}^{x}>0$.
In this case, we expect $\chi^{x}_{\rm{bottom}}$ follows the Curie law at low temperatures.
On the other hand, $M_{\rm{bottom}}^{x}=0$ in the staggered state, and then $\chi^{x}_{\rm{bottom}}$ takes a finite value at low temperatures.
The main interest in the zero-field transverse-susceptibility in the present fully frustrated model is to study which behavior is realized in the temperature dependence of $\chi_{\rm{bottom}}^{x}$.
As for $\chi^{x}_{\rm{tip}}$, the low-temperature Curie law is expected because of $M_{\rm{tip}}^{x}>0$.

In our calculation of the zero-field susceptibilities $\chi_{\rm{tip}}^{x}$ and $\chi_{\rm{bottom}}^{x}$, 
we use the notation $\Gamma_{1}=\Gamma$ and $\Gamma_{2}=\lambda\Gamma$ and write our Hamiltonian as
\begin{eqnarray}
   H=H_{0}-\Gamma Q\;\;\;\;(Q\equiv Q_1+\lambda Q_2),
\label{eq:1_1}
\end{eqnarray}
where
\begin{eqnarray}
   H_{0}=J\sum_{i=0}^{N-1}[s_{2i}^{z}s_{2i+2}^{z}+s_{2i+1}^{z}(s_{2i}^{z}+s_{2i+2}^{z})]
\label{eq:1_2}
\end{eqnarray}
is the unperturbed Hamiltonian.
Because $\lim_{\Gamma\to 0}H=H_{0}$ has a diagonal form, the calculation of transverse susceptibilities
can be made by a direct extension of Minami's method.\cite{rf:9}
In $\S2$ we show both $\chi_{\rm{tip}}^{x}$ and $\chi_{\rm{bottom}}^{x}$ diverge for $T \to 0$.

The divergence of $\chi_{\rm{tip}}^{x}$ is reasonable as mentioned previously,
but the divergence of $\chi_{\rm{bottom}}^{x}$ is not trivial.
In order to get further insight into the low temperature behavior of $\chi_{\rm{bottom}}^{x}$, 
we keep $\Gamma_{1}$ finite and study the susceptibility $\chi_{\rm{bottom}}^{x}(\Gamma_{1})$.
In this case, we write our Hamiltonian as
\begin{eqnarray}
   H=\mathcal{H}_{0}-\Gamma_{2}Q_{2},
\label{eq:1_4}
\end{eqnarray}
where
\begin{eqnarray}
   \mathcal{H}_{0}=J\sum_{i=0}^{N-1}[s_{2i}^{z}s_{2i+2}^{z}+s_{2i+1}^{z}(s_{2i}^{z}+s_{2i+2}^{z})]-\Gamma_{1}Q_{1}.
\label{eq:1_5}
\end{eqnarray}
It should be noted that $\mathcal{H}_{0}$ has off-diagonal terms, so we have to start with the diagonalization of $\mathcal{H}_{0}$. 
This procedure is explained in $\S3$.
As a result, we find that the above-mentioned divergence is suppressed by $\Gamma_{1}>0$.

Needles to say, exact solutions of one-dimensional quantum spin systems have provided testing grounds for the development of numerical calculation methods.\cite{Otsuka,Okunishi} 
A typical example is the quantum $XY$ model with nearest neighbor interactions.\cite{rf:2} 
In most of such systems, however, the structure of energy spectrum is so simple that a numerical calculation method reproduces the exact results rather easily.
Therefore, it is desired to find exact solutions of frustrated systems with complex energy spectra.
Our exact transverse susceptibility of the fully frustrated Ising $\Delta$-chain may be useful from this point of view.

The present paper is organized as follows. 
In $\S2$, we calculate $\chi_{\rm{tip}}^{x}$ and $\chi_{\rm{bottom}}^{x}$ by using Minami's transfer matrix method. 
In $\S3$, the calculation of $\chi_{\rm{bottom}}^{x}(\Gamma_{1})$ is presented, where we use an unitary transformation introduced by one of the present authors.\cite{rf:11} 
In $\S4$, we discuss the behavior of the zero-temperature susceptibility as a function of $\Gamma_{1}$ on the basis of the second-order perturbation theory. 
The results obtained in this paper are summarized in $\S5$.

\section{Exact calculation of zero-field susceptibilities $\chi_{\rm{tip}}^{x}$ and $\chi_{\rm{bottom}}^{x}$}
\subsection{Formulation of the transverse susceptibility}
The transverse susceptibility $\chi^{x}$ for the system described by the Hamiltonian (\ref{eq:1_1}) is defined as follows:
\begin{eqnarray}
   \chi^{x}&=&\frac{\partial}{\partial\Gamma}\frac{\operatorname{Tr}Qe^{-\beta H}}{\operatorname{Tr}e^{-\beta H}}\bigg|_{\Gamma =0}
\nonumber\\
   &=&\frac{\partial}{\partial\Gamma}\biggl[\frac{\Braket{Q}_{0}+\Gamma\int_{0}^{\beta}dt\Braket{e^{tH_{0}}Q
   e^{-tH_{0}}Q}_{0}}{1+\Gamma\int_{0}^{\beta}dt\Braket{Q}_{0}}+\mathcal{O}(\Gamma^{2})\biggr]\bigg|_{\Gamma =0}
\nonumber\\
   &=&\int_{0}^{\beta}dt\Braket{Q(t)Q}_{0},
\label{eq:2_1}
\end{eqnarray}
where $\beta=1/T$ is the inverse of temperature, 
$\Braket{\cdots}_{0}\equiv \operatorname{Tr}\cdots e^{-\beta H_{0}}/\operatorname{Tr}e^{-\beta H_{0}}$, 
and $Q(t)\equiv e^{tH_{0}}Qe^{-tH_{0}}$. In Eq.~(\ref{eq:2_1}), 
we have used the well known formula, 
\begin{equation}
   e^{-\beta H}\simeq e^{-\beta H_{0}}\left\{1+\Gamma\int_{0}^{\beta}dt\;Q(t)+\cdots\right\},
\label{eq:2_2}
\end{equation}
and the fact of $\Braket{Q}_{0}=0$. Then the transverse susceptibility for the antiferromagnetic Ising $\Delta$-chain is given by 
\begin{equation}
   \chi^{x}=\chi^{x}_{\rm{tip}}+\lambda^{2}\chi^{x}_{\rm{bottom}}
\label{eq:2_3}
\end{equation}
with
\begin{eqnarray}
   \chi^{x}_{\rm{tip}}&=&\int_{0}^{\beta}dt \Braket{Q_1(t)Q_1}_{0}=
   N\int_{0}^{\beta}dt\Braket{s_{1}^{x}(t)s_{1}^{x}}_{0},
\label{eq:2_4}\\
   \chi^{x}_{\rm{bottom}}&=&\int_{0}^{\beta}dt \Braket{Q_2(t)Q_2}_{0}=N\int_{0}^{\beta}dt\Braket{s_{0}^{x}(t)s_{0}^{x}}_{0},
\label{eq:2_5}
\end{eqnarray}
where we have used the translational invariance. 

\subsection{Transfer matrix}
Now we introduce the transfer matrix for Eq.~(\ref{eq:1_2}).
Let $\Ket{l}$ be the eigenstate of $s^{z}$ with the eigenvalue $d_{l}$ ($d_{1}=-d_{2}=1/2$). 
Then the transfer matrix $V$ can be written as follows:
\begin{equation}
   V_{m,n}=\sum_{l=1}^{2}V'_{m,l,n}\;\;\;\;
   \left(V'_{m,l,n}\equiv e^{-\beta J[d_{m}d_{n}+d_{l}(d_{m}+d_{n})]}\right).
\label{eq:2_6}
\end{equation}
In a matrix form, $V$ is given by
\begin{equation}
   V=\begin{pmatrix} 2e^{-\beta J/4}\cosh(\beta J/2) & 2e^{\beta J/4} \\ 2e^{\beta J/4} & 2e^{-\beta J/4}\cosh(\beta J/2) \end{pmatrix}.
\label{eq:2_7}
\end{equation}
\noindent
The maximum eigenvalue of $V$ is $\lambda_{1}=e^{-3\beta J/4}(1+3e^{\beta J})$, and the minimum one is $\lambda_{2}=e^{-3\beta J/4}(1-e^{\beta J})$.
Eigenvectors of $V$ are obtained as follows:
\begin{eqnarray}
   \bm{u}_{1}\equiv\begin{pmatrix} u_{11}\\ u_{21}\end{pmatrix}=\frac{1}{\sqrt{2}}\begin{pmatrix} 1\\ 1\end{pmatrix}{\mbox{ for $\lambda_{1}$,}}\;\;\;\;
   \bm{u}_{2}\equiv\begin{pmatrix} u_{12}\\ u_{22}\end{pmatrix}=\frac{1}{\sqrt{2}}\begin{pmatrix} 1\\ -1\end{pmatrix}{\mbox{ for $\lambda_{2}$.}}
\label{eq:2_8}
\end{eqnarray}
\noindent
Then, we can define an unitary matrix 
\begin{equation}
   U=\begin{pmatrix} u_{11} & u_{12} \\ u_{21} & u_{22} \end{pmatrix}=\frac{1}{\sqrt{2}}\begin{pmatrix} 1& 1 \\ 1 & -1 \end{pmatrix}=U^{\dagger},
\label{eq:2_10}
\end{equation}
\noindent
which diagonalizes $V$ as $\Lambda=U^{\dagger}VU$, where $\Lambda$ is a diagonal matrix whose elements are $(\Lambda)_{ij}=\lambda_{j}\delta_{ij}$. 
\noindent

For an integer $L=\mathcal{O}(N)$, we calculate $V^{L}$ as follows:
\begin{eqnarray}
   V^{L}=U^{\dagger}\Lambda^{L}U=
   \begin{pmatrix} u_{11} & u_{21} \\ u_{21} & u_{12} \end{pmatrix}
   \begin{pmatrix} \lambda_{1}^{L}& 0 \\ 0 &\lambda_{2}^{L} \end{pmatrix}
   \begin{pmatrix} u_{11} & u_{12} \\ u_{21} & u_{22} \end{pmatrix}
   \rightarrow \frac{\lambda_{1}^{L}}{2}\begin{pmatrix} 1 & 1 \\ 1 & 1 \end{pmatrix}
\label{eq:2_11}
\end{eqnarray}
in the thermodynamic limit, $N\rightarrow\infty$. 
The partition function $Z_{0}$ is
\begin{equation}
Z_{0}=\operatorname{Tr}e^{-\beta H_{0}}=\operatorname{Tr}V^{N}=\operatorname{Tr}\Lambda^{N}\rightarrow\lambda_{1}^{N}.\label{eq:2_12}
\end{equation}

\subsection{Exact expressions for the zero-field susceptibilities}
To obtain $\chi_{\rm{tip}}^{x}$ and $\chi_{\rm{bottom}}^{x}$, we need to calculate correlation functions, 
$\Braket{s_{0}^{x}(t)s_{0}^{x}}_{0}$ and $\Braket{s_{1}^{x}(t)s_{1}^{x}}_{0}$. 
Defining $h_{0}^{z}(i)\equiv J[s_{2i}^{z}s_{2i+2}^{z}+s_{2i+1}^{z}(s_{2i}^{z}+s_{2i+2}^{z})]$, 
we rewrite $\Braket{s_{1}^{x}(t)s_{1}^{x}}_{0}$ as follows:
\begin{eqnarray}
   \Braket{s_{1}^{x}(t)s_{1}^{x}}_{0}
   &=&\frac{1}{Z_{0}}\sum_{\set{l}}\Bra{\set{l}}e^{t h_{0}^{z}(0)}s_{1}^{x}
   e^{-t h_{0}^{z}(0)}s_{1}^{x}e^{-\beta H_{0}}\Ket{\set{l}}
\nonumber\\
   &=&\frac{1}{Z_{0}}\sum_{p,q=1}^{2}\sum_{l,l'=1}^{2}
   e^{t[\ce_{pq}(l)-\ce_{pq}(l')]}
   \bigl|\Braket{l|s^{x}|l'}\bigr|^{2}V'_{p,l,q}V^{N-1}_{q,p},
\label{eq:2_13}
\end{eqnarray}
where $\Ket{\set{l}}\equiv\Ket{l_{0},\cdots,l_{2N-1}}$, $\sum_{\set{l}}\equiv\sum_{l_{0}=1}^{2}\cdots\sum_{l_{2N-1}=1}^{2}$, 
and $\ce_{pq}(l)=J(d_{p}+d_{q})d_{l}$. Similarly, we obtain 
\begin{eqnarray}
   \Braket{s_{0}^{x}(t)s_{0}^{x}}_{0}
   &=&\frac{1}{Z_{0}}\sum_{\set{l}}\Bra{\set{l}}e^{t [h_{0}^{z}(-1)+ h_{0}^{z}(0)]}s_{0}^{x}
   e^{-t[ h_{0}^{z}(-1)+h_{0}^{z}(0)]}s_{0}^{x}e^{-\beta H_{0}}\Ket{\set{l}}
\nonumber\\
   &=&\frac{1}{Z_{0}}\hspace{-1mm}\sum_{p,q,r,s=1}^{2}\sum_{l,l'=1}^{2}
   \hspace{-1mm}e^{t[\ce_{pqrs}(l)-\ce_{pqrs}(l')]}
   \bigl|\Braket{l|s^{x}|l'}\bigr|^{2}V'_{p,q,l}V'_{l,r,s}V^{N-2}_{s,p}\;\;
\label{eq:2_14}
\end{eqnarray}
with $\ce_{pqrs}(l)=J(d_{p}+d_{q}+d_{r}+d_{s})d_{l}$.
The transverse susceptibilities per site, $\chi^{x}_{0,{\rm{tip}}}=\lim_{N\to\infty}\chi^{x}_{\rm{tip}}/N$ 
and $\chi^{x}_{0,{\rm{bottom}}}=\lim_{N\to\infty}\chi^{x}_{\rm{bottom}}/N$, turn out to be
\begin{eqnarray}
   \chi^{x}_{0,{\rm{tip}}}=
   \frac{1}{4\lambda_{1}}
   \sum_{p,q}\biggl[\sum_{\substack{l\\ \Delta\ce_{pq}\neq 0}}\hspace{-2mm}
   \frac{e^{-\beta\ce_{pq}(l)}}{\Delta\ce_{pq}(l)}
   +\beta\hspace{-3mm}\sum_{\substack{ l\\ \Delta\ce_{pq}=0}}\hspace{-2mm}
   e^{-\beta\ce_{pq}(l)}\biggr]e^{-\beta Jd_{p}d_{q}},
\label{eq:2_15}
\end{eqnarray}
where $\Delta\ce_{pq}(l)=\ce_{pq}(\bar{l})-\ce_{pq}(l)$ with $\bar{l}=3-l$, and
\begin{eqnarray}
   \chi^{x}_{0,{\rm{bottom}}}=
   \frac{1}{4\lambda_{1}^{2}}
   \sum_{p,q,r,s}\biggl[
   \hspace{-1mm}\sum_{\substack{l\\ \Delta\ce_{pqrs}\neq0 }} \hspace{-3mm}
  \frac{e^{-\beta\ce_{pqrs}(l)}}{\Delta\ce_{pqrs}(l)}
   +\beta\hspace{-3mm}\sum_{\substack{l\\ \Delta\ce_{pqrs}=0}}\hspace{-4mm}
   e^{-\beta\ce_{pqrs}(l)}\biggr]e^{-\beta J(d_{p}d_{q}+d_{r}d_{s})},\;\;\;\;
\label{eq:2_16}
\end{eqnarray}
where $\Delta\ce_{pqrs}(l)=\ce_{pqrs}(\bar{l})-\ce_{pqrs}(l)$.

\begin{figure}[b]
\centerline{\includegraphics[width=8cm,clip]{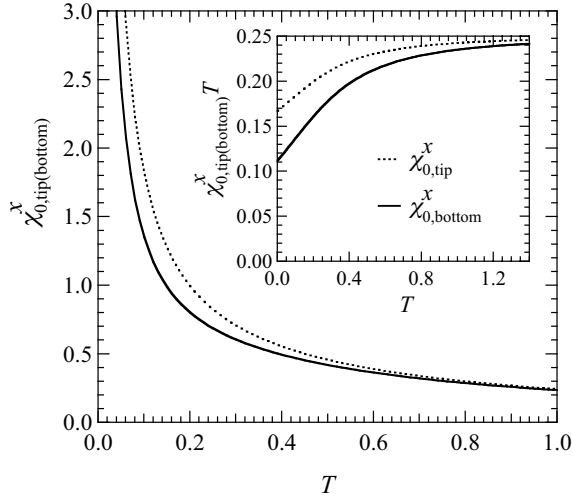}}
\caption{The exact transverse susceptibility for the antiferromagnetic Ising $\Delta$-chain. 
Equations (\ref{eq:2_17}) and (\ref{eq:2_18}) are plotted as a function of $T$.}
\label{fig:2}
\end{figure}

We carry out summations in Eqs.~(\ref{eq:2_15}) and (\ref{eq:2_16}), and then obtain the exact transverse susceptibility of tip spins and the bottom spin chain as follows:
\begin{eqnarray}
   J\chi^{x}_{0,{\rm{tip}}}&=&\frac{1}{2\lambda_{1}}\biggl[e^{\beta J/4}-e^{-3\beta J/4}+\beta Je^{\beta J/4}\biggr],\label{eq:2_17}\\
   J\chi^{x}_{0,{\rm{bottom}}}&=&\frac{1}{4\lambda_{1}^{2}}
   \biggl[9e^{\beta J/2}-8e^{-\beta J/2}-e^{-3\beta J/2}+
   2\beta J\left(e^{-\beta J/2}+2e^{\beta J/2}\right)\biggr],\;\;\;
\label{eq:2_18}
\end{eqnarray}
which are plotted in Fig.~\ref{fig:2}.
When temperature decreases, $\chi^{x}_{0,{\rm{tip}}}$ increases and diverges as expected.
The low-temperature asymptotic form of $\chi^{x}_{0,{\rm{tip}}}$ is given by
\begin{equation}
   \chi^{x}_{0,{\rm{tip}}}\simeq\frac{1}{6T}+\frac{1}{6J}.
\label{eq:2_19}
\end{equation}
The Curie constant is $1/6$, which is comparable to the value in the high-temperature limit, $1/4$.
(See the inset in Fig.~\ref{fig:2}, where $\chi^{x}_{0,{\rm{tip}}}T$ is shown.)
We also find that $\chi^{x}_{0,{\rm{bottom}}}$ diverges as $T\rightarrow 0$. 
Looking into the expression of $\chi^{x}_{0,{\rm{bottom}}}$ in Eq.~(\ref{eq:2_5}),
then we notice that it is not affected by the transverse field $\Gamma_1$ at all.
Therefore, the corresponding ground state in the calculation of $\chi^{x}_{0,{\rm{bottom}}}$
is the transverse-field paramagnetic state shown in Fig.~{\ref{fig:1b}} (b), and thus the low-temperature Curie behavior
appears in $\chi^{x}_{0,{\rm{bottom}}}$ due to a finite bottom-spin magnetization along the $x$ direction induced by the infinitely small $\Gamma_2$ at zero temperature.
The asymptotic form of $\chi^{x}_{0,{\rm{bottom}}}$ is
\begin{equation}
   \chi^{x}_{0,{\rm{bottom}}}\simeq\frac{1}{9T}+\frac{1}{4J}.
\label{eq:2_20}
\end{equation}
The Curie constant $1/9$ is somewhat smaller than that of $\chi^{x}_{0,{\rm{tip}}}$.
The temperature dependence of $\chi^{x}_{0,{\rm{bottom}}}T$ is shown in the inset in Fig.~\ref{fig:2}

In the above susceptibility calculations, we have set $\Gamma_1,\;\Gamma_2\rightarrow 0$ at the same time.
If we keep $\Gamma_1$ finite and calculate $\chi^{x}_{0,{\rm{bottom}}}(\Gamma_1)$, then the corresponding ground state
turns to the staggered state shown in Fig.~{\ref{fig:1b}} (a) and different low-temperature behavior is expected.
In order to study this issue, we carry out an exact calculation of $\chi^{x}_{0,{\rm{bottom}}}(\Gamma_1)$ in the following section.

\section{Exact calculation of transverse susceptibility $\chi_{\rm{bottom}}^{x}(\Gamma_{1})$}

\subsection{Unitary transformation of $\mathcal{H}_{0}$}
In this sectoin, we consider the Hamiltonian in Eq.~(\ref{eq:1_4}) for finite values of $\Gamma_{1}$. 
In order to carry out the exact calculation of the transverse susceptibility, we have to diagonalize $\mathcal{H}_{0}$. 
The diagonalization is achieved by introducing the following operator:
\begin{equation}
   R_{i}=\sum_{\alpha,\beta=1}^{2}P_{i}(d_{\alpha},d_{\beta})e^{i2\theta(d_{\alpha},d_{\beta})s_{2i+1}^{y}},
\label{eq:3_1}
\end{equation}
where 
\begin{equation}
   P_{i}(d_{\alpha},d_{\beta})=\frac{1}{4}(1+4d_{\alpha}s_{2i}^{z})(1+4d_{\beta}s_{2i+2}^{z})
\label{eq:3_2}
\end{equation}
is the projection operator for the two bottom spins on the $i$th triangle. The angle $\theta$ is defined by 
\begin{eqnarray}
   \theta(d_{1},d_{1})=-\theta(d_{2},d_{2})=\tan^{-1}\frac{\Omega-J}{\Gamma_{1}},\;\;\;
   \theta(d_{1},d_{2})=-\theta(d_{2},d_{1})=\frac{\pi}{4},
\label{eq:3_4}
\end{eqnarray}
where $\Omega=\sqrt{J^{2}+\Gamma_{1}^{2}}$.
This angle represents rotation angle of the tip spin depending on the bottom spin state.\cite{rf:11}
(See Appendix for the derivation of $R_{i}$.)
Then, the unitary transformation $R$ diagonalizing Eq.~(\ref{eq:1_5}) is 
\begin{equation}
   R=\prod_{i=0}^{N-1}R_{i}.\label{eq:3_5}
\end{equation}
Carrying out the unitary transformation by $R$, we obtain the transformed Hamiltonian $\tilde{H}$ as follows:
\begin{eqnarray}
\tilde{H}&=&R^{\dagger}HR=\tilde{\mathcal{H}}_{0}-\Gamma_{2}\tilde{Q}_{2},\label{eq:3_6}\\
\tilde{\mathcal{H}}_{0}&=&R^{\dagger}\mathcal{H}_{0}R=J\sum_{i=0}^{N-1}s_{2i}^{z}s_{2i+2}^{z}+\sum_{i=0}^{N-1}\left(J_{+}s_{2i}^{z}+J_{-}s_{2i+2}^{z}\right)s_{2i+1}^{z},\label{eq:3_7}\\
\tilde{Q}_{2}&=&R^{\dagger}Q_{2}R=\sum_{i=0}^{N-1}\left[I_{0}s_{2i}^{x}+I_{1}s_{2i}^{x}s_{2i-1}^{y}s_{2i+1}^{y}+I_{+}s_{2i}^{y}s_{2i+1}^{y}-I_{-}s_{2i}^{y}s_{2i-1}^{y}\right],\label{eq:3_8}
\end{eqnarray}
where the coefficients are defined by
\begin{eqnarray}
   J_{\pm}=\Omega\pm\Gamma_{1},\;\;I_{0}=\frac{J}{2\Omega},\;\;I_{1}=\frac{2J}{\Omega},\;\;I_{\pm}=\frac{\Omega\pm\Gamma_{1}}{\Omega}.
\label{eq:3_9}
\end{eqnarray}

\subsection{Transfer matrix}

Let us introduce the transfer matrix $\mathcal{V}$ for $\tilde{\mathcal{H}}_{0}$.
As the same way in $\S2$, it is given by
\begin{equation}
   \mathcal{V}_{m,n}=\sum_{l=1}^{2}\mathcal{V}'_{m,l,n}\;\;\;\;
   \left(\mathcal{V}'_{m,l,n}\equiv e^{-\beta[Jd_{m}d_{n}+\left(J_{+}d_{m}+J_{-}d_{n}\right)d_{l}]}\right),
\label{eq:3_12}
\end{equation}
or, in a matrix form,
\begin{equation}
   \mathcal{V}=\begin{pmatrix}2e^{-\beta J/4}\cosh\beta \Omega & 2e^{\beta J/4}\cosh\beta\Gamma_{1}\\2e^{\beta J/4}\cosh\beta\Gamma_{1} & 2e^{-\beta J/4}\cosh\beta \Omega\end{pmatrix}.
\label{eq:3_13}
\end{equation}
The maximam and minimum eigenvalues of $\mathcal{V}$ are, respectively, 
$\tilde{\lambda}_{1}=2e^{-\beta J/4}\cosh\beta \Omega+2e^{\beta J/4}\cosh\beta\Gamma_{1}$ 
and $\tilde{\lambda}_{2}=2e^{-\beta J/4}\cosh\beta \Omega-2e^{\beta J/4}\cosh\beta\Gamma_{1}$.
The corresponding eigenvectors are those in Eq.~(\ref{eq:2_8}). The partition function is
\begin{equation}
   \tilde{Z}_{0}=\operatorname{Tr}e^{-\beta \tilde{\mathcal{H}}_{0}}=\operatorname{Tr}\mathcal{V}^{N}\rightarrow\tilde{\lambda}_{1}^{N}.
\end{equation}

\subsection{Exact expression for $\chi^{x}_{\rm{bottom}}(\Gamma_{1})$}

The transverse susceptibility for $\tilde{H}$ in Eq.~(\ref{eq:3_6}) can be written as
\begin{equation}
   \chi^{x}_{\rm{bottom}}(\Gamma_{1})=\int_{0}^{\beta}dt\Braket{\tilde{Q}_{2}(t)\tilde{Q}_{2}}_{0},\label{eq:3_14}
\end{equation}
where 
\begin{eqnarray}
   \Braket{\tilde{Q}_{2}(t)\tilde{Q}_{2}}_{0}&=&N\Bigg[I_{0}^{2}\Braket{s_{0}^{x}(t)s_{0}^{x}}_{0}+I_{1}^{2}
   \Braket{[s_{0}^{x}s_{-1}^{y}s_{1}^{y}](t)s_{0}^{x}s_{-1}^{y}s_{1}^{y}}_{0}\nonumber\\
   &&\hspace{6mm}+I_{+}^{2}\Braket{[s_{0}^{y}s_{1}^{y}](t)s_{0}^{y}s_{1}^{y}}_{0}+I_{-}^{2}\Braket{[s_{0}^{y}s_{-1}^{y}](t)s_{0}^{y}s_{-1}^{y}}_{0}\Bigg].
\label{eq:3_15}
\end{eqnarray}
In the above equations, we have redefined as $\Braket{\cdots}_{0}\equiv \operatorname{Tr}\cdots e^{-\beta \tilde{\mathcal{H}}_{0}}/\operatorname{Tr}e^{-\beta \tilde{\mathcal{H}}_{0}}$,
$\tilde{Q}_{2}(t)\equiv e^{t\tilde{\mathcal{H}}_{0}}\tilde{Q}_{2}e^{-t\tilde{\mathcal{H}}_{0}}$ and so on.
In the thermodynamic limit $N\to\infty$, we obtain 
\begin{eqnarray}
   \chi^{x}_{0,\rm{bottom}}(\Gamma_{1})&\equiv&\lim_{N\to\infty}\chi_{\rm{bottom}}^{x}(\Gamma_{1})/N
\nonumber\\
   &=&\lim_{N\to\infty}\int_{0}^{\beta}dt\bigg[I_{0}^{2}\Braket{s_{0}^{x}(t)s_{0}^{x}}_{0}+I_{1}^{2}
   \Braket{\left[s_{0}^{x}s_{-1}^{y}s_{1}^{y}\right](t)s_{0}^{x}s_{-1}^{y}s_{1}^{y}}_{0}
\nonumber\\
   &&\hspace{19mm}+I_{+}^{2}\Braket{\left[s_{0}^{y}s_{1}^{y}\right](t)s_{0}^{y}s_{1}^{y}}_{0}+I_{-}^{2}
   \Braket{\left[s_{0}^{y}s_{-1}^{y}\right](t)s_{0}^{y}s_{-1}^{y}}_{0}\bigg].\;\;\;\;\;\;
\label{eq:3_16}
\end{eqnarray}
\noindent

It is convenient to define $\tilde{h}_{0}^{z}(i)\equiv Js_{2i}^{z}s_{2i+2}^{z}+(J_{+}s_{2i}^{z}+J_{-}s_{2i+2}^{z})s_{2i+1}^{z}$.
Then, the correlation functions are written as follows:
\begin{eqnarray}
   \Braket{s_{0}^{x}(t)s_{0}^{x}}_{0}
   &=&\frac{1}{\tilde{Z}_{0}}\sum_{\set{l}}\Bra{\set{l}}e^{t\left[\tilde{h}_{0}^{z}(-1)+\tilde{h}_{0}^{z}(0)\right]}s_{0}^{x}
   e^{-t\left[\tilde{h}_{0}^{z}(-1)+\tilde{h}_{0}^{z}(0)\right]}s_{0}^{x}e^{-\beta \tilde{H_{0}}}\Ket{\set{l}}
\nonumber\\
   &=&\frac{1}{\tilde{Z}_{0}}\sum_{p,q,r,s}\sum_{l_{0},l_{0}'}e^{t\left[\tilde{\mathcal{E}}_{pqrs}(l_{0})-
   \tilde{\mathcal{E}}_{pqrs}(l_{0}')\right]}\left|\Braket{l_{0}|s_{0}^{x}|l_{0}'}\right|^{2}
   \mathcal{V}'_{p,q,l_{0}}\mathcal{V}'_{l_{0},r,s}\mathcal{V}^{N-2}_{s,p}\;\;\;\;\;\;
\label{eq:3_17}
\end{eqnarray}
with $\tilde{\mathcal{E}}_{pqrs}(l_{0})=J(d_{p}+d_{s})d_{l_{0}}+J_{+}d_{r}d_{l_{0}}+J_{-}d_{q}d_{l_{0}}$,
\begin{eqnarray}
   &&\Braket{[s_{0}^{x}s_{-1}^{y}s_{1}^{y}](t)s_{0}^{x}s_{-1}^{y}s_{1}^{y}}_{0}
\nonumber\\
   &&\hspace{10mm}=\frac{1}{\tilde{Z}_{0}}\sum_{\set{l}}\Bra{\set{l}}e^{t\left[\tilde{h}_{0}^{z}(-1)+\tilde{h}_{0}^{z}(0)\right]}s_{0}^{x}s_{-1}^{y}s_{1}^{y}
   e^{-t\left[\tilde{h}_{0}^{z}(-1)+\tilde{h}_{0}^{z}(0)\right]}s_{0}^{x}s_{-1}^{y}s_{1}^{y}e^{-\beta \tilde{H_{0}}}\Ket{\set{l}}
\nonumber\\
   &&\hspace{10mm}=\frac{1}{\tilde{Z}_{0}}\sum_{p,q}\sum_{\substack{l_{0},l_{1},l_{-1}\\l_{0}',l_{1}',l_{-1}'}}
   e^{t\left[\tilde{\mathcal{E}}_{pq}(l_{0},l_{-1},l_{1})-\tilde{\mathcal{E}}_{pq}(l_{0}',l_{-1}',l_{1}')\right]}\left|\Braket{l_{0}|s_{0}^{x}|l_{0}'}\right|^{2}
\nonumber\\
   &&\hspace{35mm}\times\left|\Braket{l_{-1}|s_{-1}^{y}|l_{-1}'}\right|^{2}\left|\Braket{l_{1}|s_{1}^{y}|l_{1}'}\right|^{2}
   \mathcal{V}'_{q,l_{-1},l_{0}}\mathcal{V}'_{l_{0},l_{1},p}\mathcal{V}^{N-2}_{p,q}\hspace{10mm}
\label{eq:3_18}
\end{eqnarray}
with $\tilde{\mathcal{E}}_{pq}(l_{0},l_{-1},l_{1})=J(d_{p}+d_{q})d_{l_{0}}+J_{+}(d_{l_{0}}d_{l_{1}}+d_{q}d_{l_{-1}})+J_{-}(d_{l_{1}}d_{p}+d_{l_{0}}d_{l_{-1}})$,
\begin{eqnarray}
&&\Braket{[s_{0}^{y}s_{1}^{y}](t)s_{0}^{y}s_{1}^{y}}_{0}
\nonumber\\
   &&\hspace{5mm}=\frac{1}{\tilde{Z}_{0}}\sum_{\set{l}}\Bra{\set{l}}e^{t\left[\tilde{h}_{0}^{z}(-1)+\tilde{h}_{0}^{z}(0)\right]}s_{0}^{y}s_{1}^{y}
   e^{-t\left[\tilde{h}_{0}^{z}(-1)+\tilde{h}_{0}^{z}(0)\right]}s_{0}^{y}s_{1}^{y}e^{-\beta \tilde{H_{0}}}\Ket{\set{l}}
\nonumber\\
   &&\hspace{5mm}=\frac{1}{\tilde{Z}_{0}}\sum_{p,q,r}\sum_{\substack{l_{0},l_{1}\\l_{0}',l_{1}'}}
   e^{t\left[\tilde{\mathcal{E}}^+_{pqr}(l_{0},l_{1})-\tilde{\mathcal{E}}^+_{pqr}(l_{0}',l_{1}')\right]}
   \left|\Braket{l_{0}|s_{0}^{y}|l_{0}'}\right|^{2}\left|\Braket{l_{1}|s_{1}^{y}|l_{1}'}\right|^{2}\mathcal{V}'_{p,q,l_{0}}\mathcal{V}'_{l_{0},l_{1},r}\mathcal{V}^{N-2}_{r,p}
\nonumber\\
\label{eq:3_19}
\end{eqnarray}
with 
$\tilde{\mathcal{E}}^+_{pqr}(l_{0},l_{1})=J(d_{p}+d_{r})d_{l_{0}}+J_{+}d_{l_{0}}d_{l_{1}}+J_{-}(d_{l_{1}}d_{p}+d_{l_{0}}d_{q})$,
and
\begin{eqnarray}
   &&\Braket{[s_{0}^{y}s_{-1}^{y}](t)s_{0}^{y}s_{-1}^{y}}_{0}
\nonumber\\
   &&\hspace{8mm}=\frac{1}{\tilde{Z}_{0}}\sum_{\set{l}}\Bra{\set{l}}e^{t\left[\tilde{h}_{0}^{z}(-1)+\tilde{h}_{0}^{z}(0)\right]}s_{0}^{y}s_{-1}^{y}
   e^{-t\left[\tilde{h}_{0}^{z}(-1)+\tilde{h}_{0}^{z}(0)\right]}s_{0}^{y}s_{-1}^{y}e^{-\beta \tilde{H_{0}}}\Ket{\set{l}}
\nonumber\\
   &&\hspace{8mm}=\frac{1}{\tilde{Z}_{0}}\sum_{p,q,r}\sum_{\substack{l_{0},l_{-1}\\l_{0}',l_{-1}'}}
   \hspace{-2mm}e^{t\left[\tilde{\mathcal{E}}^-_{pqr}(l_{0},l_{-1})-\tilde{\mathcal{E}}^-_{pqr}(l_{0}',l_{-1}')\right]}
   \left|\Braket{l_{0}|s_{0}^{y}|l_{0}'}\right|^{2}\left|\Braket{l_{-1}|s_{-1}^{y}|l_{-1}'}\right|^{2}
\nonumber\\
   &&\hspace{85mm}\times 
   \mathcal{V}'_{p,l_{-1},l_{0}}\mathcal{V}'_{l_{0},q,r}\mathcal{V}^{N-2}_{r,p}\;\;\;\;
\label{eq:3_20}
\end{eqnarray}
with $\tilde{\mathcal{E}}^-_{pqr}(l_{0},l_{-1})=J(d_{p}+d_{r})d_{l_{0}}+J_{+}(d_{p}d_{l_{-1}}+d_{l_{0}}d_{q})+J_{-}d_{l_{0}}d_{l_{-1}}$.

We calculate summations in Eqs.~(\ref{eq:3_17})-(\ref{eq:3_20}), and substitute those results into Eq.~(\ref{eq:3_16}). 
Carrying out the integration over $t$, we obtain our final expression:
\begin{eqnarray}
   \chi_{0,\rm{bottom}}^{x}(\Gamma_{1})=I_{0}^{2}C_{xx}+I_{1}^{2}C_{xyy}+I_{+}^{2}C_{yy}^++I_{-}^{2}C_{yy}^-.
\label{eq:3_21}
\end{eqnarray}
Here, $C_{xx}$ is given by
\begin{eqnarray}
   C_{xx}&=&\lim_{N\to\infty}\int_{0}^{\beta}dt\Braket{s_{0}^{x}(t)s_{0}^{x}}_{0}
\nonumber\\
   &=&\frac{1}{4\lambda_{1}^{2}}\sum_{p,q,r,s}
   \frac{e^{-\beta\tilde{\mathcal{E}}_{pqrs}(2)}-e^{-\beta\tilde{\mathcal{E}}_{pqrs}(1)}}{\tilde{\mathcal{E}}_{pqrs}(1)-\tilde{\mathcal{E}}_{pqrs}(2)}
   e^{-\beta (J_{+}d_{p}d_{q}+J_{-}d_{r}d_{s})}
\nonumber\\
   &=&
   \frac{1}{2\lambda_1'}\bigg[
   e^{-\beta\Gamma_{1}}(f_{J+\Omega}+f_{J-\Omega})+e^{-2\beta\Gamma_{1}}f_{J-\Gamma_1}+f_{J+\Gamma_1}
\nonumber\\
   &&\hspace{7mm}
   +2e^{-\beta(J-\Omega+2\Gamma_1)/2}f_\Omega\cosh\frac{\beta\Gamma_1}{2}
   +2e^{-\beta(J+\Gamma_1)/2}f_{\Gamma_1}\cosh\frac{\beta \Omega}{2}
   \bigg],\;\;\;\;
\label{eq:3_22}
\end{eqnarray}
where $\lambda_{1}'=[1+e^{-\beta\Gamma_{1}}+2e^{-\beta(J+\Gamma_{1})/2}\cosh\beta \Omega]^{2}$
and $f_x=(1-e^{-\beta x})/x$.
Similarly, $C_{xyy}$ denotes
\begin{eqnarray}
   C_{xyy}&=&\lim_{N\to\infty}\int_{0}^{\beta}dt\Braket{[s_{0}^{x}s_{-1}^{y}s_{1}^{y}](t)s_{0}^{x}s_{-1}^{y}s_{1}^{y}}_{0}
\nonumber\\
   &=&\frac{1}{64\lambda_{1}^{2}}
   \sum_{p,q}\sum_{l_{0},l_{-1},l_{1}}\hspace{-1mm}
   \frac{e^{-\beta\tilde{\mathcal{E}}_{pq}(l_{0},l_{-1},l_{1})}}
   {\tilde{\mathcal{E}}_{pq}(\bar{l}_{0},\bar{l}_{-1},\bar{l}_{1})-\tilde{\mathcal{E}}_{pq}(l_{0},l_{-1},l_{1})}
\nonumber\\
   &=&
   \frac{C_{xx}}{16},
\label{eq:3_23}
\end{eqnarray}
where we have used the notation $\bar{l}_i=3-l_i$. 
Finally, $C_{yy}^{\pm}$ denote 
\begin{eqnarray}
   C_{yy}^{\pm}&=&\lim_{N\to\infty}\int_{0}^{\beta}dt\Braket{[s_{0}^{y}s_{\pm 1}^{y}](t)s_{0}^{y}s_{\pm 1}^{y}}_{0}
\nonumber\\
   &=&\frac{1}{16\lambda_{1}^{2}}
   \sum_{p,q,r}\hspace{-2mm}\sum_{\substack{l_{0},l\\\Delta\tilde{\mathcal{E}}^{\pm}_{pqr}\neq0}}\hspace{-2mm}
   \frac{e^{-\beta[\tilde{\mathcal{E}}^{\pm}_{pqr}(l_{0},l)+J_{\pm}d_{q}d_{r}]}}{\Delta\tilde{\mathcal{E}}^{\pm}_{pqr}(l_{0},l)}
   +\frac{\beta}{32\lambda_{1}^{2}}\sum_{p,q,r}\hspace{-2mm}
   \sum_{\substack{l_{0},l\\\Delta\tilde{\mathcal{E}}_{pqr}^{\pm}=0}}\hspace{-2mm}
   e^{-\beta[\tilde{\mathcal{E}}^{\pm}_{pqr}(l_{0},l)+J_{\pm}d_{q}d_{r}]}
\nonumber\\
   &=&\frac{1}{8\lambda_{1}'}
   \bigg[
   e^{-2\beta\Gamma_{1}}f_{J\pm \Omega-\Gamma_{1}}+f_{J\mp \Omega+\Gamma_{1}}+2e^{-\beta\Gamma_{1}}f_{J}
\nonumber\\
   &&\hspace{8mm}
   +2e^{-\beta(J-\Omega+2\Gamma_{1}\pm\Gamma_1)/2}f_{\Omega\mp\Gamma_{1}}
   +2\beta e^{-\beta(J+2\Gamma_{1})/2}\cosh\frac{\beta(\Omega\pm\Gamma_1)}{2}
   \bigg],
\label{eq:3_24}
\end{eqnarray}
where $\Delta\tilde{\mathcal{E}}^{\pm}_{pqr}(l_{0},l)=\tilde{\mathcal{E}}^{\pm}_{pqr}(\bar{l}_{0},\bar{l})-\tilde{\mathcal{E}}^{\pm}_{pqr}(l_{0},l)$.
If we set $\Gamma_{1}\to 0$ in our final expression, we obtain
\begin{equation}
   \lim_{\Gamma_{1}\to 0}J\chi_{0,\rm{bottom}}^{x}(\Gamma_{1})
   =\frac{5-4e^{-\beta J}-e^{-2\beta J}+\beta J(2+e^{-\beta J})}{2(3+e^{-\beta J})^{2}},
\label{eq:3_26}
\end{equation}
which is the same as $\chi_{0,\rm{bottom}}^{x}$ obtained in the previous section.

\begin{figure}[hbt]
\begin{center}
\includegraphics[width=12cm,clip]{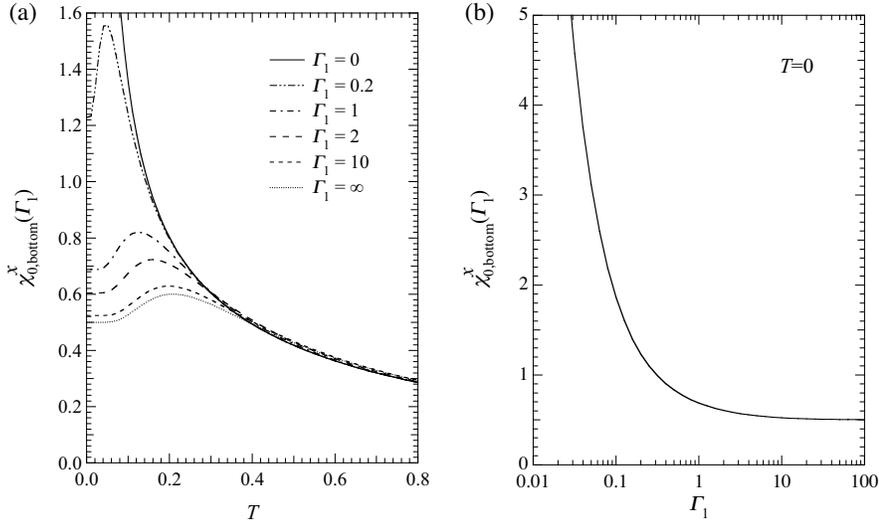}
\end{center}
\caption{(a) Temperature dependence of $\chi_{0,\rm{bottom}}^{x}(\Gamma_{1})$
for several values of $\Gamma_{1}$, and (b) zero-temperature value of
$\chi_{0,\rm{bottom}}^{x}(\Gamma_{1})$ as a function of $\Gamma_{1}$,
where we set $J=1$.}
\label{fig:3}
\end{figure}

In Fig.~\ref{fig:3} (a), we show the temperature dependence of $\chi_{0,\rm{bottom}}^{x}(\Gamma_1)$ for $\Gamma_{1}=0, 0.2, 1, 2, 10$ and $\infty$ in the unit of $J=1$.
It is found that $\chi_{0,\rm{bottom}}^{x}(\Gamma_1>0)$ has a plateau at low temperatures instead of the Curie law of $\chi_{0,\rm{bottom}}^{x}(\Gamma_1=0)$.
The appearance of such plateau structure is naturally expected if the staggered ground state is realized.
The most transparent situation in this context is the large $\Gamma_1$ limit. 
Setting $\Gamma_1\rightarrow\infty$ in our expression of $\chi_{0,\rm{bottom}}^{x}(\Gamma_1)$, we obtain
\begin{equation}
   \lim_{\Gamma_{1}\to\infty}J\chi_{0,\rm{bottom}}^{x}(\Gamma_{1})=
   \frac{1}{2}\left[\frac{\beta J}{\cosh^2 (\beta J/4)}+\tanh (\beta J/4)\right],
\label{eq:3_26b}
\end{equation}
which is nothing but the transverse-field susceptibility of the antiferromagnetic Ising chain with nearest neighbor interactions.
In the limit $\Gamma_1\rightarrow\infty$, the tip spins polarize to the $x$ direction completely and the remaining degrees of freedom are just bottom spins,
which follow the nearest-neighbor Ising model.
For the nearest-neighbor Ising chain, it has been well known that such a plateau structure appears in the temperature dependence of the transverse susceptibility.\cite{rf:9}

It is interesting to study the zero-temperature value of $\chi_{0,\rm{bottom}}^{x}(\Gamma_1)$.
Assuming $\Gamma_{1}\neq 0$, we take the limit of $T\to 0$ in Eq.~(\ref{eq:3_21}). Then we get
\begin{eqnarray}
   \lim_{T\to 0}\chi_{0,\rm{bottom}}^{x}(\Gamma_{1})
   =\frac{1}{8}\left(\frac{4}{J}+\frac{1}{\Gamma_{1}}+\frac{1}{J+\Gamma_{1}}\right),
\label{eq:3_28}
\end{eqnarray}
which is plotted as a function of $\Gamma_1$ in Fig.~\ref{fig:3} (b).
When $\Gamma_1\rightarrow 0$, the zero-temperature susceptibility diverges like $1/8\Gamma_1$.
On the other hand, it approaches to $1/2J$ for large $\Gamma_1$.

\section{Zero-temperature transverse susceptibility: perturbation study}
In the previous section we have obtained the exact expression of the transverse susceptibility $\chi_{0,\rm{bottom}}^{x}(\Gamma_{1})$
and found the low-temperature plateau structure for $\Gamma_{1}>0$. 
In order to get deeper understanding to this behavior, we here use the second-order perturbation theory to calculate the zero-temperature transverse susceptibility.

\begin{figure}[hbt]
\begin{center}
   \includegraphics[width=10cm,clip]{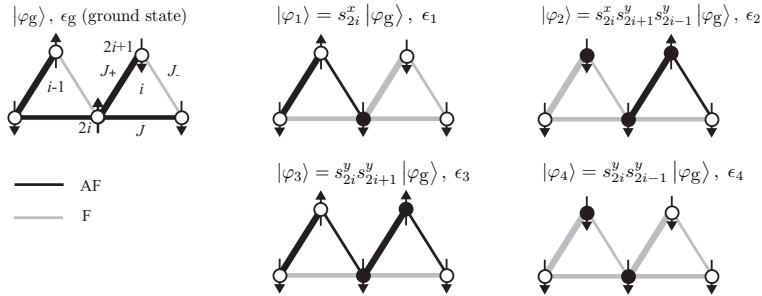}
\end{center}
\caption{Spin configurations of the ground state $\Ket{\varphi_{\mbox{g}}}$ with energy 
$\epsilon_{\mbox{g}}$ and intermediate states $\Ket{\varphi_{i}}$ with the energy 
$\epsilon_{i}$ ($i=1\sim 4$) in the second-order perturbation theory.
The strength of the exchanges is $J_{+}>J>J_{-}$.
The black circles represent spins flipped by a perturbation operator.
The black and gray bonds denote antiferromagnetic (AF) and ferromagnetic (F) bonds, respectively.
}
\label{fig:perturb}
\end{figure}

The macroscopic degeneracy of the Ising $\Delta$-chain is removed by $\Gamma_{1}$, and we obtain the unique ground state $\Ket{\varphi_{\mbox{g}}}$ of $\tilde{\mathcal{H}}_0$ as shown in Fig.~\ref{fig:perturb},
which is, of course, equivalent to the staggered state.
The ground-state energy is
\begin{equation}
   \epsilon_{\mbox{g}}=-\frac{N}{2}\left(J+J_{+}-J_{-}\right).
   \label{eq:4_1}
\end{equation}
The perturbation operator $\tilde{Q}_2$ in Eq.~(\ref{eq:3_8}) is applied to this ground state,
then we obtain four types of intermediate states, $\Ket{\varphi_{i}}$ ($i=1\sim 4$), with the energy of $\epsilon_{i}$, which are defined in Fig.~\ref{fig:perturb}.
The energy denominators, $\Delta E_{i}\equiv\epsilon_{i}-\epsilon_{\mbox{g}}$, are
\begin{eqnarray}
\Delta E_{1}=\Delta E_{2}=J+\frac{J_{+}}{2}-\frac{J_{-}}{2},\;\;
\Delta E_{3}=J-\frac{J_{-}}{2},\;\;
\Delta E_{4}=J+\frac{J_{+}}{2}.
\label{eq:4_2}
\end{eqnarray}
Thus, ground-state energy $E_{\mbox{g}}(\Gamma_{2})$ perturbed by the transverse field $\Gamma_2$ is given by
\begin{eqnarray}
   E_{\mbox{g}}(\Gamma_{2})=\epsilon_{\mbox{g}}-N\left(\frac{\Gamma_{2}}{\Omega}\right)^{2}
   \left[\frac{J^{2}}{8(J+\Gamma_{1})}+\frac{(\Omega+\Gamma_{1})^{2}}{16(J-\Omega+\Gamma_{1})}+\frac{(\Omega-\Gamma_{1})^{2}}{16(J+\Omega+\Gamma_{1})}\right].\;\;\;\;
\label{eq:4_5}
\end{eqnarray}
The zero-temperature susceptibility is obtained by
\begin{eqnarray}
   \chi_{0,\rm{bottom}}^{x}(\Gamma_{1})\bigg|_{T=0}=-\frac{\partial^{2}}{\partial\Gamma_{2}^{2}}\left(\frac{E_{\mbox{g}}}{N}\right),
\label{eq:4_6}
\end{eqnarray}
which leads to Eq.~(\ref{eq:3_28}).
Let us consider the limit $\Gamma_{1}\to0$, where $\chi_{0,\rm{bottom}}^{x}(\Gamma_{1})\simeq 1/8\Gamma_1$. 
Noting $\Delta E_{3}\simeq\mathcal{O}(\Gamma_{1})$, 
we find that the divergent behavior for $\Gamma_{1}\to0$ stems from the intermediate state $\Ket{\varphi_{3}}$,
which is a member of the degenerate ground-state manifold of the Ising $\Delta$-chain with no transverse fields.
(Note that the number of antiferromagnetic bonds in $\Ket{\varphi_{3}}$ is the same as $\Ket{\varphi_{\mbox{g}}}$.)
In other words, introduction of finite tip-spin field $\Gamma_1$ opens 
an energy gap between $\Ket{\varphi_{\mbox{g}}}$ and $\Ket{\varphi_{3}}$
and thus prevents the divergence of $\chi_{0,\rm{bottom}}^{x}(\Gamma_{1})$ at low temperatures $T<\sim\Gamma_1$. 

\section{Summary}
In this paper, we have studied the transverse susceptibility of the antiferromagnetic Ising $\Delta$-chain.
The spin-$1/2$ antiferromagnetic Ising $\Delta$-chain has next-nearest-neighbor interactions, so it can not be mapped to a solvable model by the Jordan-Wigner transformation.
We have employed Minami's method to carry out an exact calculation of the transverse susceptibility of the Ising $\Delta$-chain.

First, we have considered the system on which infinitesimal fields $\Gamma_{1}$ on tip spins and $\Gamma_{2}$ on bottom spins are applied.
Using the transfer-matrix method, we have obtained the transverse susceptibilities, $\chi^{x}_{0,\rm{tip}}$ for $\Gamma_{1}$ and $\chi^{x}_{\rm{0,bottom}}$ for $\Gamma_{2}$. 
We have found that both $\chi^{x}_{0,\rm{tip}}$ and $\chi^{x}_{\rm{0,bottom}}$ shows Curie type behavior at low temperatures.

Next, we have extended our calculation to the the antiferromagnetic Ising $\Delta$-chain under a finite field $\Gamma_{1}$ on tip spins and an infinitesimal field $\Gamma_{2}$ on the bottom spin chain. 
We have diagonalized the Hamiltonian by an unitary operator $R$. 
Introducing the transfer matrix for this transformed Hamiltonian $\tilde{\mathcal{H}}_{0}$, we have carried out the exact calculation of the transverse susceptibility for $\Gamma_{2}$ which is denoted by $\chi_{0,\rm{bottom}}^{x}(\Gamma_{1})$.  
We have found that finite $\Gamma_1$ changes the low-temperature behavior and $\chi_{0,\rm{bottom}}^{x}(\Gamma_{1})$ shows the low-temperature plateau.
The role of $\Gamma_1$ has been also discussed by using the second-order perturbation theory, by which the spin configurations
giving rise to the low-temperature Curie law have been identified.

\section*{Acknowledgement}
We would like to thank Professor A. Oguchi for valuable comments.

\appendix
\section{} 
%
In this Appendix, we describe how to introduce the operator $R_{i}$ in Eq.~(\ref{eq:3_1}). Let us rewrite the Hamiltonian Eq.~(\ref{eq:1_5}) as follows:
\begin{eqnarray}
   \mathcal{H}_{0}=J\sum_{i=0}^{N-1}s_{2i}^{z}s_{2i+2}^{z}+\sum_{i=0}^{N-1}h_{i}(s_{2i}^{z},s_{2i+2}^{z}),
\label{eq:A_2}
\end{eqnarray}
where 
\begin{eqnarray}
   h_{i}(s_{2i}^{z},s_{2i+2}^{z})&=&Js_{2i+1}^{z}(s_{2i}^{z}+s_{2i+2}^{z})-\Gamma_{1}s_{2i+1}^{x}\nonumber\\
   &=&\frac{1}{2}\begin{pmatrix}J(s_{2i}^{z}+s_{2i+2}^{z})&-\Gamma_{1}\\-\Gamma_{1}&-J(s_{2i}^{z}+s_{2i+2}^{z})\end{pmatrix}.
\label{eq:A_4}
\end{eqnarray}
\noindent
In the second line of Eq.~(\ref{eq:A_4}), we have used the matrix representations of $s_{2i+1}^{z}$ and $s_{2i+1}^{x}$.

When $s_{2i}^{z}=s_{2i+2}^{z}=1/2$, the eigenvalue equation for the matrix in Eq.~(\ref{eq:A_4}) is
\begin{equation}
   \begin{pmatrix}
   J & -\Gamma_{1}\\-\Gamma_{1} & -J\end{pmatrix}\begin{pmatrix}a\\b\end{pmatrix}=2\epsilon\begin{pmatrix}a\\b
   \end{pmatrix}.
\label{eq:A_5}
\end{equation}
Eigenvalues are calculated as
\begin{equation}
   \epsilon=\pm\frac{1}{2}\sqrt{J^{2}+\Gamma_{1}^{2}}\equiv\pm \frac{\Omega}{2},
\label{eq:A_6}
\end{equation}
and eigenvectors are
\begin{eqnarray}
   &&\begin{pmatrix}a\\b\end{pmatrix}=\frac{1}{\sqrt{\Gamma_{1}^{2}+(\Omega-J)^{2}}}
   \begin{pmatrix}
   \Gamma_{1}\\\mp(\Omega-J)\end{pmatrix}=\begin{pmatrix}\cos\theta\\\mp\sin\theta
   \end{pmatrix}\mbox{ for $\epsilon=\pm \frac{\Omega}{2}$}.
\label{eq:A_7a}
\end{eqnarray}
From Eq.~(\ref{eq:A_7a}), we can obtain an unitary matrix $R_{i}(d_{1},d_{1})$ which diagonalizes $h_{i}(d_{1},d_{1})$:
\begin{equation}
   R_{i}(d_{1},d_{1})=
   \begin{pmatrix}\cos\theta & \sin\theta\\-\sin\theta & \cos\theta\end{pmatrix}
   =\cos\theta+i2s_{2i+1}^{y}\sin\theta=e^{2i\theta s_{2i+1}^{y}}.
\label{eq:A_8}
\end{equation}
In the same way, we can also obtain the unitary matrix $R_{i}(d_{2},d_{2})$ for $s_{2i}^{z}=s_{2i+2}^{z}=-1/2$. 
The eigenvalue equation
\begin{eqnarray}
   \begin{pmatrix}-J & -\Gamma_{1}\\-\Gamma_{1} & J\end{pmatrix}
   \begin{pmatrix}a\\b\end{pmatrix}=2\epsilon\begin{pmatrix}a\\b\end{pmatrix},
\label{eq:A_9a}
\end{eqnarray}
leads to the following eigenvalues and eigenvectors:
\begin{subequations}
\begin{eqnarray}
   a=-\sin\theta,\; b=\cos\theta\;\;\;\mbox{for}\;\epsilon=\frac{\Omega}{2},
\label{eq:A_9b}
\end{eqnarray}
\begin{eqnarray}
   a=\cos\theta,\;b=\sin\theta\;\;\;\mbox{for}\;\epsilon=-\frac{\Omega}{2}.
\label{eq:A_9c}
\end{eqnarray}
\end{subequations}
Thus, the unitary matrix $R_{i}(d_{2},d_{2})$ diagonalizing $h_{i}(d_{2},d_{2})$ is
\begin{equation}
   R_{i}(d_{2},d_{2})=\begin{pmatrix}\cos\theta & -\sin\theta\\\sin\theta & \cos\theta\end{pmatrix}=e^{-2i\theta s_{2i+1}^{y}}.
\label{eq:A_10}
\end{equation}
\noindent
For the case of $s_{2i}^{z}=-s_{2i+2}^{z}$, the eigenvalue equation
\begin{eqnarray}
   \begin{pmatrix}0 & -\Gamma_{1}\\-\Gamma_{1} & 0\end{pmatrix}
   \begin{pmatrix}a\\b\end{pmatrix}=2\epsilon\begin{pmatrix}a\\b\end{pmatrix},
\label{eq:A_11a}
\end{eqnarray}
gives
\begin{subequations}
\begin{eqnarray}
   a=-b=\frac{1}{\sqrt{2}}\;\;\;\mbox{for}\;\epsilon=\frac{\Gamma_{1}}{2},
\label{eq:A_11b}
\end{eqnarray}
\begin{eqnarray}
   a=b=\frac{1}{\sqrt{2}}\;\;\;\mbox{for}\;\epsilon=-\frac{\Gamma_{1}}{2}.
\label{eq:A_11c}
\end{eqnarray}
\end{subequations}
Then $R_{i}(d_{1},d_{2})$ and $R_{i}(d_{2},d_{1})$ are, respectively, given by
\begin{equation}
   R_{i}(d_{1},d_{2})=\frac{1}{\sqrt{2}}\begin{pmatrix}1 & 1\\-1 & 1\end{pmatrix}=e^{i\frac{\pi}{2}s_{2i+1}^{y}},
\label{eq:A_12}
\end{equation}
\begin{equation}
   R_{i}(d_{2},d_{1})=\frac{1}{\sqrt{2}}\begin{pmatrix}1 & -1\\1 & 1\end{pmatrix}=e^{-i\frac{\pi}{2}s_{2i+1}^{y}}.
\label{eq:A_13}
\end{equation}

From Eqs.~(\ref{eq:A_8}), (\ref{eq:A_10}), (\ref{eq:A_12}) and (\ref{eq:A_13}), 
we can define $R_{i}$ using the projection operator (\ref{eq:3_2}) as follows:
\begin{equation}
   R_{i}=\sum_{\alpha=1}^{2}\sum_{\beta=1}^{2}P_{i}(d_{\alpha},d_{\beta})R_{i}(d_{\alpha},d_{\beta}),
\label{eq:A_14}
\end{equation}
which is Eq.~(\ref{eq:3_1}).


\begin{thebibliography}{99}
  
\bibitem{rf:1} B.K. Chakrabarti, A. Das, and J. Inoue, Eur. Phys. J. \andvol{B9,1999,233}.
\bibitem{rf:2} S. Katsura, Phys.~Rev. \andvol{127,1962,1508}.
\bibitem{rf:3} P. Pfeuty, Ann. Phys. \andvol{57,1970,79}.
\bibitem{rf:4} M. E. Fisher, Physica \andvol{26,1960,618}.
\bibitem{rf:5} M. E. Fisher, J. Math. Phys. \andvol{4,1963,124}.
\bibitem{rf:6} T. Idogaki, M. Rikitoku, and J. W. Tucker, J. Magn. Magn. Mater. \andvol{152,1996,311}.
\bibitem{rf:7} T. Kaneyoshi, \PTP{98,1997,57}.
\bibitem{rf:8} K. Minami, \JP{A29,1996,6395}.
\bibitem{rf:9} K. Minami, \JPSJ{67,1998,2255}.
\bibitem{rf:10} R. Moessner and S. L. Sondhi, \PRB{63,2001,224401}.
\bibitem{rf:11} Y. Fukumoto and A. Oguchi, \JPSJ{72,2003,2317}.
\bibitem{Otsuka} H. Otsuka, Phys. Rev. B \PRB{51,1995,305}.
\bibitem{Okunishi} K. Okunishi, Phys. Rev. B \PRB{60,1999,4043}.
\end{thebibliography}
\end{document}